\documentclass[runningheads]{llncs}
\usepackage[T1]{fontenc}
\usepackage{graphicx}
\usepackage{multirow} 
\usepackage{makecell}
\usepackage{booktabs}
\usepackage{amsfonts}
\usepackage{ulem}
\usepackage{marvosym}
\usepackage{url}
\usepackage{color}
\usepackage[marginal]{footmisc}
\usepackage{hyperref}
\makeatletter
\def\UrlAlphabet{%
      \do\a\do\b\do\c\do\d\do\e\do\f\do\g\do\h\do\i\do\j%
      \do\k\do\l\do\m\do\n\do\o\do\p\do\q\do\r\do\s\do\t%
      \do\u\do\v\do\w\do\x\do\y\do\z\do\A\do\B\do\C\do\D%
      \do\E\do\F\do\G\do\H\do\I\do\J\do\K\do\L\do\M\do\N%
      \do\O\do\P\do\Q\do\R\do\S\do\T\do\U\do\V\do\W\do\X%
      \do\Y\do\Z}
\def\UrlDigits{\do\1\do\2\do\3\do\4\do\5\do\6\do\7\do\8\do\9\do\0}
\g@addto@macro{\UrlBreaks}{\UrlOrds}
\g@addto@macro{\UrlBreaks}{\UrlAlphabet}
\g@addto@macro{\UrlBreaks}{\UrlDigits}
\makeatother

\def\ie{\normalem\emph{i.e.}}

\begin{document}
%
% \title{Conventional Model Fusion with Effective Post-processing for Stroke Lesion Segmentation}
% \titlerunning{Conventional Model Fusion with Effective Post-processing}

\title{MAPPING: Model Average with Post-processing for Stroke Lesion Segmentation}
\titlerunning{MAPPING}

% \title{SEE: Stroke Lesion Segmentation using Ensemble Model and Effective Post-processing}
% \titlerunning{SEE}

\author{Jiayu Huo\inst{1}\thanks{Equal contribution}\textsuperscript{(\Letter)} \and
Liyun Chen\inst{2}\inst{*} \and
Yang Liu\inst{1}\inst{*} \and
Maxence Boels\inst{1} \and
Alejandro Granados\inst{1} \and
S\'{e}bastien Ourselin\inst{1} \and
Rachel Sparks\inst{1}
}

\authorrunning{J. Huo et al.}

\institute{School of Biomedical Engineering and Imaging Sciences (BMEIS), King's College London, London, UK \\
\email{jiayu.huo@kcl.ac.uk} \\
% \email{rachel.sparks@kcl.ac.uk} \\
\and
School of Biomedical Engineering, Shanghai Jiao Tong University, Shanghai, China \\
}

% \author{Paper ID: None}

\maketitle              % typeset the header of the contribution

% \footnote{*J. Huo, L. Chen and Y. Liu contribute equally to this work.\\}

\begin{abstract}
Accurate stroke lesion segmentation plays a pivotal role in stroke rehabilitation research, to provide lesion shape and size information which can be used for quantification of the extent of the stroke and to assess treatment efficacy. Recently, automatic segmentation algorithms using deep learning techniques have been developed and achieved promising results. In this report, we present our stroke lesion segmentation model based on nnU-Net framework, and apply it to the Anatomical Tracings of Lesions After Stroke (ATLAS v2.0) dataset. Furthermore, we describe an effective post-processing strategy that can improve some segmentation metrics. Our method took the first place in the 2022 MICCAI ATLAS Challenge with an average Dice score of 0.6667, Lesion-wise F1 score of 0.5643, Simple Lesion Count score of 4.5367, and Volume Difference score of 8804.9102. Our code and trained model weights are publicly available at \url{https://github.com/King-HAW/ATLAS-R2-Docker-Submission}.

\end{abstract}

\section{Introduction}
Strokes are one of the most common neurosurgical conditions that can cause disability, as well as being in the top three causes of death worldwide \cite{kim2020global}. Magnetic resonance imaging (MRI) is often utilized in clinical practice to provide a clear representation of soft brain tissues and lesion areas. Accurate stroke lesion mask is needed for the purpose of lesion localization and quantification, as it can measure the overlap between lesion areas and brain structures to further relate human behavior changes with brain abnormalities \cite{boyd2017biomarkers,cassidy2018neuroimaging}. Currently, stroke lesion segmentation achieved by expert clinicians remains the golden standard for lesion quantification and evaluation. However, voxel-level manual labelling is a tedious and time-consuming task. More importantly, it requires domain knowledge expertise and results should be validated by two or more neurologists. In order to mitigate the burden of manual lesion delineation, stroke MRI datasets with the voxel-wise annotation of lesion area have been released for building accurate automatic stroke lesion segmentation models.

The ATLAS v1.2 dataset \cite{liew2018large}, a stroke dataset that contains 229 subjects, was released in 2018 to provide an benchmark to evaluate stroke lesion segmentation methods with the ultimate aim of improving the performance of such methods. With the development of deep learning techniques especially the convolutional neural network, many stroke lesion segmentation methods have been presented. Most model architectures are based on U-Net \cite{ronneberger2015u} (an encoder-decoder-based model design with skip connections), and some modifications have been made to improve the segmentation performance, such as residual connections \cite{tomita2020automatic}, an auxiliary adversarial branch \cite{wang2022brain} and multi-channel inputs \cite{zhang2020mi}. However, the evaluation metrics reported in the above-mentioned manuscripts may be inflated because there is no hidden test set for ATLAS v1.2. Therefore, researchers can overfit their model by applying improper validation \cite{liew2022large} or hyperparameters may not be generalisable to unseen data.

\begin{figure}[!t]
\centering
\includegraphics[width=0.95\textwidth]{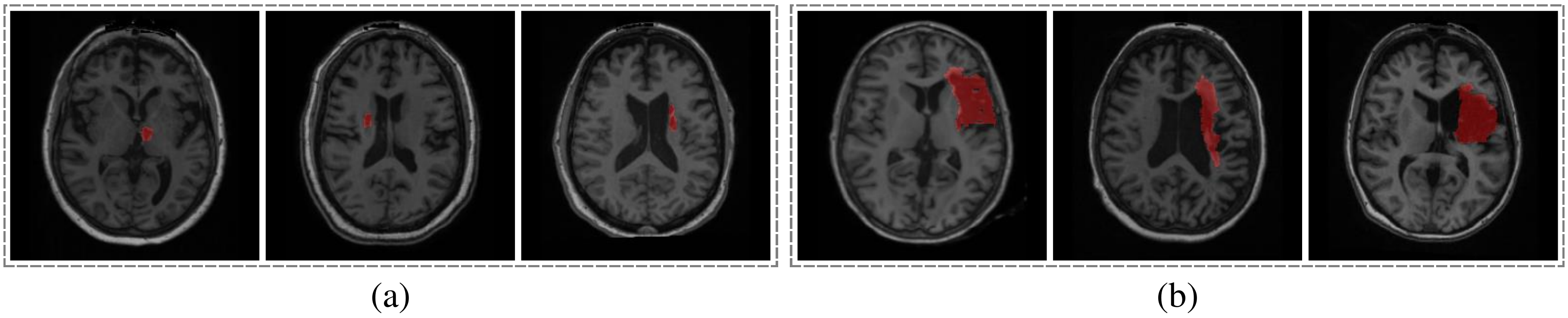}
\caption{(a) Exemplar images of small lesions. (b) Exemplar images of big lesions. Exemplar images indicate the large variance of stroke lesion shapes and sizes in ATLAS v2.0 dataset.}
\label{fig:exemplar_lesions}
\end{figure}

The ATLAS v2.0 \cite{liew2022large} is a new dataset built upon ATLAS v1.2 to address the aforementioned concerns. ATLAS v2.0 contains 1271 T1-weighted MRIs with manually annotated lesion masks collected from 44 different research cohorts across 11 countries worldwide. Furthermore, it splits all subjects into three sets: the training set, the public test set, and the hidden test set. Participants can use the training set for model design and parameter tuning, and submit the label predictions of the public test set and docker image for the hidden test set via the online evaluation platform\footnote{\url{https://atlas.grand-challenge.org/}\label{atlas_website}} to evaluate their models. Here, we show some exemplar images contained in the training set in Fig. \ref{fig:exemplar_lesions}, and we can see the large variance of stroke lesion shapes and sizes in the set. Additionally, ATLAS v2.0 is the dataset for MICCAI 2022 Stroke Segmentation Challenge \textsuperscript{\ref{atlas_website}}. 18 teams participated in this challenge and 7 teams submitted their final solutions.

In this manuscript, we describe a stroke lesion segmentation model to improve the ability of the model to generalize and prevent overfitting. This model took the first place in ATLAS v2.0 challenge for MICCAI 2022. Specifically, our model is based on nnU-Net architecture \cite{isensee2021nnu} using four different training schemes to increase the model diversity. An ensemble technique is applied to fuse predictions across training schemes to give better lesion segmentations. Additionally, a post-processing strategy is presented to further improve some evaluation metrics. Experiments on the training and test sets demonstrates the efficacy of this approach.

\section{Method}
\begin{figure}[!t]
\centering
\includegraphics[width=0.95\textwidth]{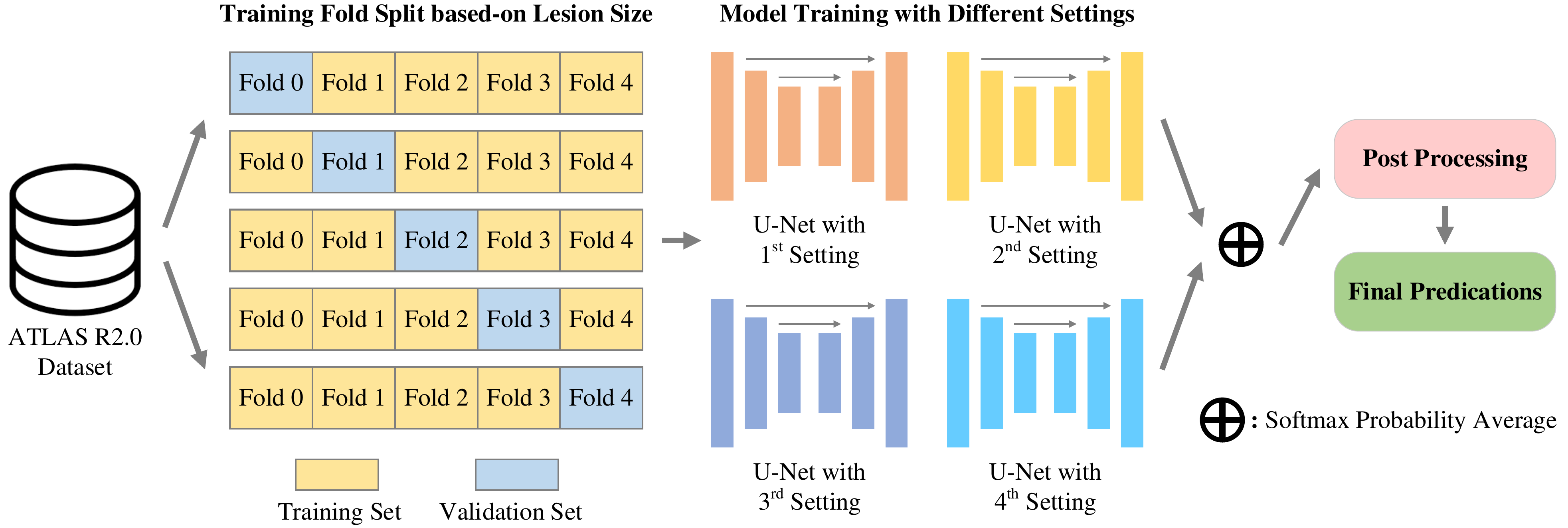}
\caption{The pipeline of our framework for ATLAS R2.0 Stroke Lesion Segmentation Challenge.} 
\label{fig:framework}
\end{figure}

The pipeline of our method is illustrated in Fig. \ref{fig:framework}. The five fold split of the training set stratified by lesion size is described in Section \ref{section_size_balanced_cross_validation}. This fold split can make sure that the local training set and validation set have similar lesion size distribution. The model training with different settings and model fusion is described in Section \ref{section_conventional_model_fusion}. We use four different schemes for segmentation model training to generate diverse predictions, which is beneficial for the model ensemble. The effective post-processing strategy is described in Section \ref{section_effective_postprocessing_strategy} which can be used to further improve some of the evaluation metrics.

\subsection{Size-balanced Cross-validation}
\label{section_size_balanced_cross_validation}
The ATLAS v2.0 dataset encompasses three parts: (1) a training set, which consists of 655 T1-weighted MRIs and lesion segmentation masks derived from 33 research cohorts, (2) a test set, which contains 300 T1-weighted MRIs collected from the same cohorts as the training set, but with lesion segmentation masks hidden, and (3) a holdout test set, which is comprised of 316 completely hidden T1-weighted MRIs and lesion segmentation masks from 11 independent cohorts. All images were normalized and registered to the MNI-152 template. 

\begin{figure}[!t]
\centering
\includegraphics[width=0.95\textwidth]{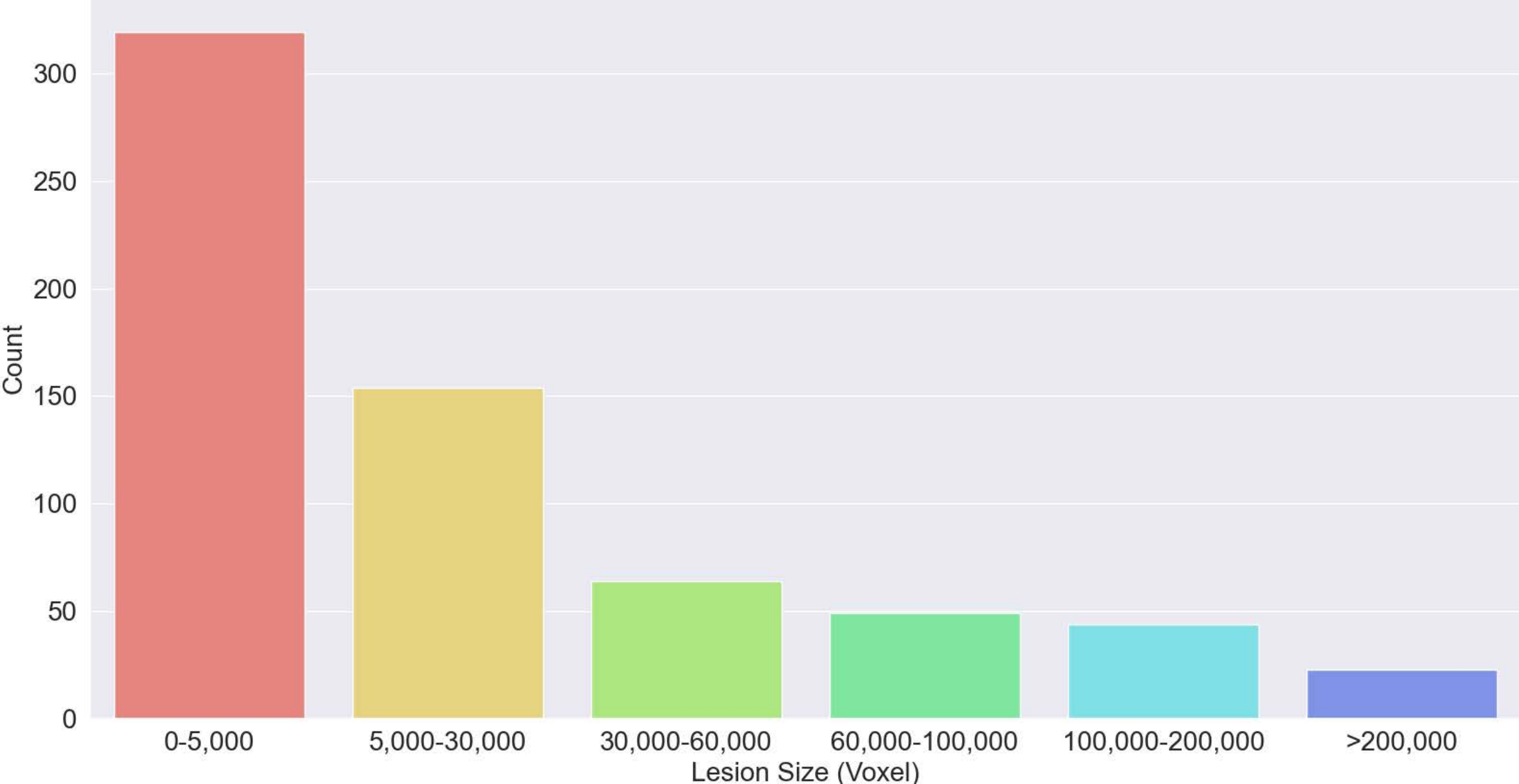}
\caption{The data distribution of the ATLAS v2.0 training set according to lesion size.}
\label{fig:lesion_distribution}
\end{figure}

We first conduct exploratory data analysis on the training set. The distribution of lesion sizes in the training set is shown in Fig. \ref{fig:lesion_distribution}. Lesion volumes ranges from 13 to 200,000+ voxels with over half of cases containing small lesions (foreground voxels $\leq$ 5,000). Therefore, we first split the training set into five folds stratified by lesion size to guarantee that the final segmentation performance will be good for across lesion size ,also this can ensure that each fold almost has the same lesion size distribution. We call this fold spilt 'size-balanced cross-validation'. Additionally, we design different training schemes to segment both small and large lesions.

\subsection{Conventional Model Fusion}
\label{section_conventional_model_fusion}

We implement all models based on nnU-Net framework \cite{isensee2021nnu}. The four different training schemes were devised so that the final predictions were more diverse. The first training scheme (Default) uses the default compound loss (Dice plus cross-entropy) to train the generic U-Net. The input image is downsampled five times in the generic U-Net. Instance normalization \cite{ulyanov2016instance} is used for feature map normalization, and Leaky ReLU \cite{maas2013rectifier} is used for nonlinear mapping. In the second training scheme (DTK10), we change the compound loss to TopK10 loss \cite{fan2017learning}. By replacing the loss function, the segmentation network is penalized when making mistakes segmenting foreground voxels, this loss further improves the segmentation performance on small lesions in particular. In the third training scheme (Res U-Net), we change the network architecture by adding a residual block \cite{he2016deep} to the generic U-Net. By adding the shortcut connections, the network structure is more complex compared with the generic U-Net which may result in diverse predictions. For the fourth training scheme (Self-Training), we use a self-training strategy to retrain the first training scheme (generic U-Net). Specifically, we generate pseudo-lesion masks for the public test set by conducting the forward inference on the best model evaluated on the original training set, and then combine these datasets with the original training dataset to train the generic U-Net from scratch.

After finishing training of four different schemes, we run the forward inference for each case through each model to obtain the probability map. Then all probability maps of each subject are averaged to attain the final probability map. Finally, the aggregated probability map is utilized as the input for our proposed post-processing strategy.

\subsection{Effective Post-processing Strategy}
\label{section_effective_postprocessing_strategy}
\begin{figure}[!t]
\centering
\includegraphics[width=0.95\textwidth]{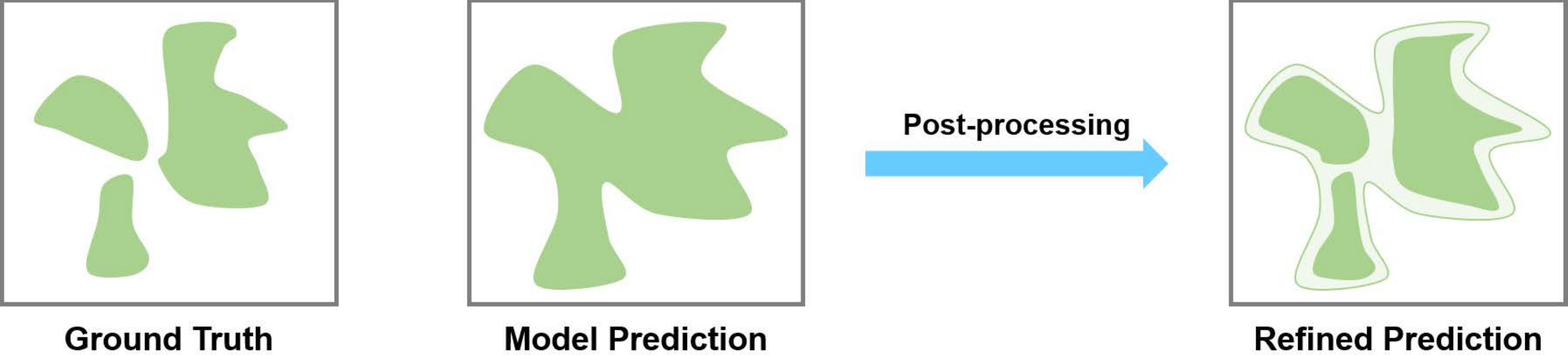}
\caption{The diagram of our proposed effective post-processing strategy.}
\label{fig:postprocessing}
\end{figure}

The post-processing strategy aims to match the number of connected components of predictions with the ground truth. The pipeline of our proposed post-processing strategy is illustrated in Fig. \ref{fig:postprocessing}. During on our exploratory data analysis we observed that the number of connected components is greater for larger lesions compared to small lesions. Therefore, our post-processing step aims to first decrease the number of connected components for small lesions by eliminating small connected components whose max probability is less than 0.7. Then, we separate the connected components into several parts by applying a high threshold (\ie, 0.55, the default threshold for getting the foreground is 0.5) for larger lesions. This post-processing strategy, increases the dice coefficient to 0.6667 from 0.6640, and the volume difference to 8805 pixels from 8891, which ranks the first on the public test set.

\section{Experimental Results}

% \subsection{Dataset}
% We utilize the ATLAS v2.0 dataset for model training and evaluation. 

\subsection{Implementation Details}
Our method is based on nnU-Net \cite{isensee2021nnu}. Image patches of $128 \times 128 \times 128$ are extracted based on the foreground lesion voxels as model input. The batch size is set to 2 and z-score normalization is applied to all images. We use the stochastic gradient descent (SGD) as the optimizer, and set the initial learning rate equal to 0.01 and momentum to 0.99. We train all models for 1000 epochs, and Kaiming initialization \cite{he2015delving} initialize all model weights.

\subsection{Evaluation Metrics}
Four metrics are utilized by the challenge organizers to evaluate the segmentation performance: Dice score (Dice), lesion-wise F1 score (L-F1) \cite{petzsche2022isles}, simple lesion count (SLC) \cite{petzsche2022isles}, and volume difference (VD). The detailed information of L-F1 and SLC can be found at this code repository\footnote{\url{https://github.com/npnl/atlas2_grand_challenge/blob/main/isles/scoring.py}}. In addition to these metric we calculate the 95\% Hausdorff distance (95HD) to evaluate the boundary errors between the model prediction and the ground-truth segmentations.

\subsection{Size-balanced Cross-validation Results}
We evaluate all training schemes, the model ensemble method, and post-processing strategy on the size-balanced cross-validation dataset. Results are shown in Table. \ref{tab:train_seg_results}. The training sets for the first three training schemes (Default, DTK10, and Res U-Net) are the same, which is the original ATLAS v2.0 training set. Res U-Net has the best Dice score among these trainig schemes, while Default achieves the best results for the remaining metrics. This indicates the nnU-Net default setting is already been able to segment the stroke lesion area accurately, as well as showing its powerful generalizability. When we generate the pseudo-lesion masks for the public test set and further combine it with the original training set for the self-training scheme, Dice and VD improve compared to the first three training schemes. This demonstrates that utilizing a semi-supervised method such as pseudo-labeling can boost model performance. The ensemble of the four training schemes gives the best results on L-F1, SLC and 95HD, while Dice and VD remain similar compared with the Self-Training scheme. Finally by adding the post-processing strategy (Ensemble (PP)) model performance can further improve the L-F1, SLC, and 95HD metrics.

\begin{table}[!t]\scriptsize
\caption{Size-balanced cross-validation results on the ATLAS v2.0 training set. Default represents the first training scheme. DTK10 represents the second training scheme. Res U-Net represents the third training scheme. Self-Training represents the fourth training scheme. Ensemble represents when the predictions of four schemes are averaged. PP indicates post-processing strategy. The best results are shown in \textcolor{red}{red} font.}
\label{tab:train_seg_results}
\centering
\linespread{1.3}\selectfont
\begin{tabular}{l|m{1.75cm}<{\centering}|m{1.75cm}<{\centering}|m{1.75cm}<{\centering}|m{1.75cm}<{\centering}|m{1.75cm}<{\centering}}
\toprule[1.5pt]

% \hline
\multirow{2}{*}{\textbf{Method}} &
\multicolumn{5}{c}{\textbf{Metrics}} \\
% \multicolumn{2}{c}{Multi-Row and Col} \\
\cline{2-6} & Dice $\uparrow$  & L-F1 $\uparrow$ & SLC $\downarrow$ & VD $\downarrow$ & 95HD $\downarrow$ \\
\hline
Default       & 0.636$\pm$0.269 & 0.549$\pm$0.282 & 3.431$\pm$6.753 & 5852$\pm$11792 & 23.297$\pm$27.598 \\
\hline
DTK10         & 0.630$\pm$0.274 & 0.547$\pm$0.291 & 3.501$\pm$6.685 & 6013$\pm$12740 & 22.360$\pm$27.501 \\
\hline
Res U-Net     & 0.639$\pm$0.265 & 0.540$\pm$0.277 & 3.562$\pm$6.631 & 5897$\pm$12034 & 23.922$\pm$28.432 \\
\hline
Self-Training & \textcolor{red}{0.649$\pm$0.261} & 0.550$\pm$0.278 & 3.530$\pm$6.636 & \textcolor{red}{5628$\pm$11084} & 24.769$\pm$29.689 \\
\hline
Ensemble      & 0.647$\pm$0.265 & 0.569$\pm$0.285 & 3.441$\pm$6.798 & 5688$\pm$11555 & 21.681$\pm$26.596 \\
\hline
Ensemble (PP) & 0.646$\pm$0.270 & \textcolor{red}{0.575$\pm$0.296} & \textcolor{red}{3.382$\pm$6.786} & 5729$\pm$11565 & \textcolor{red}{21.511$\pm$26.828} \\
% \hline

\bottomrule[1.5pt]
\end{tabular}
\end{table}

\subsection{Public Test Set Results}
We evaluate all training schemes on the public test set, and the results are shown in Table \ref{tab:test_seg_results}. Please note that L-F1 scores for the Default and DTK10 setting are missing due to modifications of the online evaluation system which occurred on 25th, June, 2022. As shown in Table \ref{tab:test_seg_results}, the performance of the four different training schemes are quite similar. By fusing all predictions together, the ensemble model can achieve the best Dice score which is $0.664\pm0.259$. By deploying the post-processing strategy, Dice can reach $0.667\pm0.259$. Ensemble (PP) also has the highest L-F1 and SLC metrics of all the evaluated models, while the VD remains comparable to Ensemble.

\begin{table}[!t]\scriptsize
\caption{Segmentation results for the different training schemes on the ATLAS v2.0 public test set. 95HD is not manually computes as lesion masks are unavailable. L-F1 scores for Default and DTK10 training schemes are missing due to the modification of the evaluation system. The best results are shown in \textcolor{red}{red} font.}
\label{tab:test_seg_results}
\centering
\linespread{1.3}\selectfont
\begin{tabular}{l|m{1.75cm}<{\centering}|m{1.75cm}<{\centering}|m{1.75cm}<{\centering}|m{1.75cm}<{\centering}}
\toprule[1.5pt]

% \hline
\multirow{2}{*}{\textbf{Method}} &
\multicolumn{4}{c}{\textbf{Metrics}} \\
% \multicolumn{2}{c}{Multi-Row and Col} \\
\cline{2-5} & Dice $\uparrow$  & L-F1 $\uparrow$ & SLC $\downarrow$ & VD $\downarrow$ \\
\hline
Default       & 0.661$\pm$0.258 &        -        & 4.670$\pm$7.021 & 8921$\pm$16996 \\
\hline
DTK10         & 0.660$\pm$0.260 &        -        & 4.617$\pm$7.026 & 8891$\pm$17079 \\
\hline
Res U-Net     & 0.659$\pm$0.258 & 0.541$\pm$0.257 & 4.547$\pm$6.758 & \textcolor{red}{8779$\pm$16895} \\
\hline
Self-Training & 0.661$\pm$0.258 & 0.545$\pm$0.258 & 4.680$\pm$6.990 & 9119$\pm$17419 \\
\hline
Ensemble      & 0.664$\pm$0.259 & 0.555$\pm$0.265 & 4.653$\pm$7.045 & 8891$\pm$17086 \\
\hline
Ensemble (PP) & \textcolor{red}{0.667$\pm$0.259} & \textcolor{red}{0.564$\pm$0.277} & \textcolor{red}{4.537$\pm$7.022} & 8805$\pm$16773 \\
% \hline

\bottomrule[1.5pt]
\end{tabular}
\end{table}

\subsection{Hidden Test Set Results}
The leaderboard of the hidden test set is provided at this link\footnote{\url{https://atlas.grand-challenge.org/evaluation/lesion-segmentation-hidden-test-set/leaderboard/}}. Our method took the first place on the leaderboard, demonstrating that our ensemble model can be further deployed to the unseen dataset and achieve superior segmentation performance to other methods. The final leaderboard rankings according to ATLAS-MICCAI score is reproduced in Table \ref{tab:test_set_rank}. The ATLAS-MICCAI score is calculated based on a weighted sum of the rankings of public test set and hidden test set. The weight of the hidden test set ranking is 4 times higher than the public test set ranking, which can reduce the overfitting risk and better evaluate the generalizability of each method. Our team (CTRL) ranks the first on the final leaderboard, demonstrating the robustness of the presented method.

\begin{table}[!t]\scriptsize
\caption{The final leaderboard of the ATLAS challenge. The ranking weight of the hidden test set is 4 times higher than the public test ranking, in order to discourage participants from overfitting on the public test data. The best results are shown in \textcolor{red}{red} font.}
\label{tab:test_set_rank}
\centering
\linespread{1.3}\selectfont
\begin{tabular}{l|m{4cm}<{\centering}|m{2cm}<{\centering}}
\toprule[1.5pt]

% \hline
\textbf{Team Name} & \textbf{ATLAS-MICCAI Score} $\downarrow$ & \textbf{Final Rank} \\
\hline
CTRL     & \textcolor{red}{3.317} & \textcolor{red}{1} \\
\hline
POBOTRI  & 3.739 & 2 \\
\hline
MIRC     & 4.937 & 3 \\
\hline
Yileinus & 5.053 & 4 \\
\hline
PLORAS   & 5.304 & 5 \\
\hline
BIA      & 6.190 & 6 \\
\hline
AICONS   & 8.080 & 7 \\
% \hline

\bottomrule[1.5pt]
\end{tabular}
\end{table}

\subsection{Visualization Results}
\begin{figure}[!t]
\centering
\includegraphics[width=0.95\textwidth]{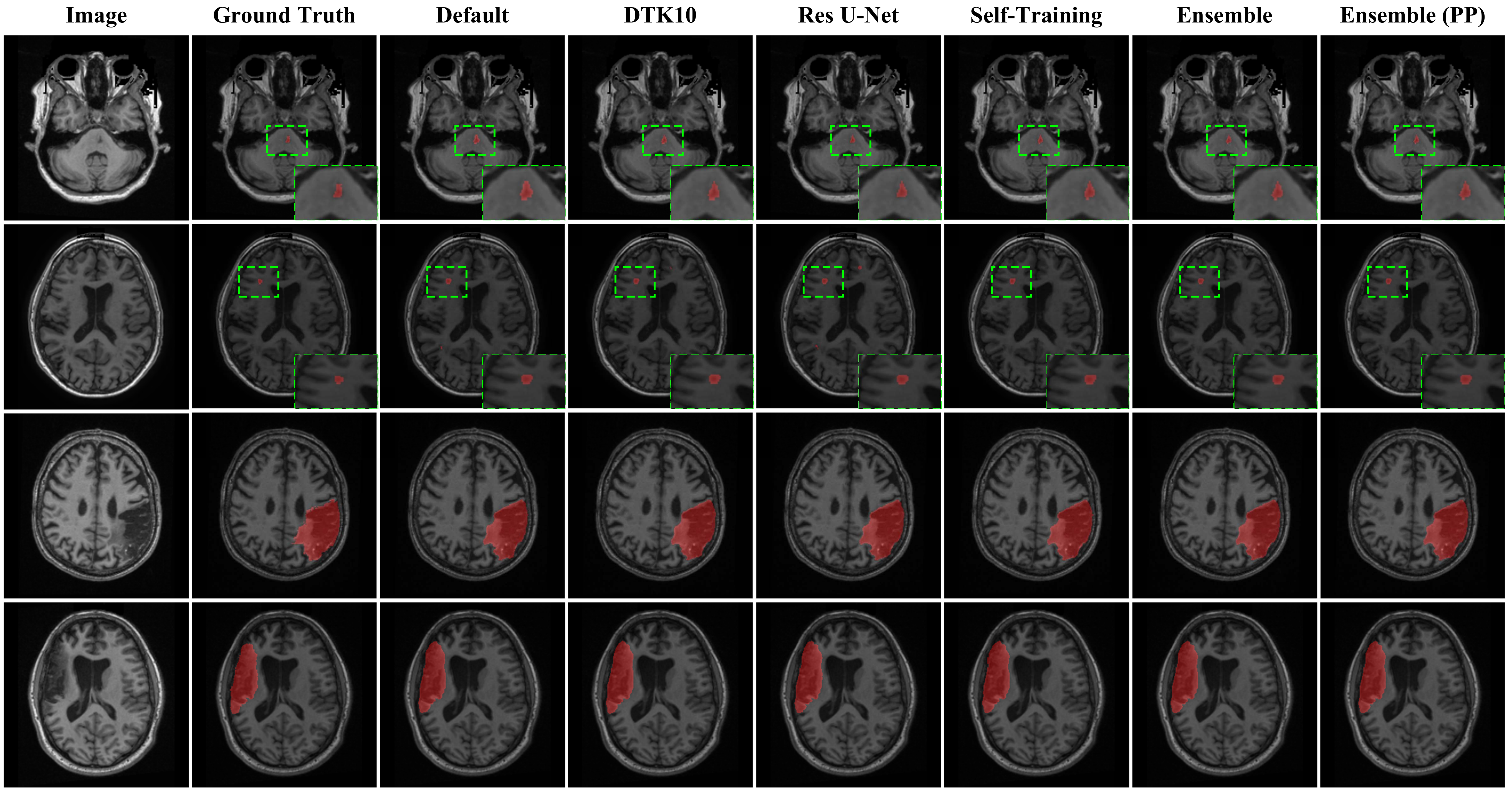}
\caption{Segmentation results for different training schemes.} 
\label{fig:seg_results}
\end{figure}

Fig. \ref{fig:seg_results} provides qualitative segmentation results for all of the training schemes. Although some of the training schemes, such as Res U-Net, may give false positive results, the overall predictions are accurate for both small and big stroke lesions. Additionally, the outputs of four different schemes (see Section \ref{section_conventional_model_fusion}) are similar with the ground truth, which means all models are able to handle the stroke lesion segmentation task. Besides, boundary areas of ensemble predictions are more smooth compared to four training schemes, indicating some noise false-positive predictions can be suppressed by averaging outputs.

\subsection{Failure Analysis}
\begin{figure}[!t]
\centering
\includegraphics[width=0.95\textwidth]{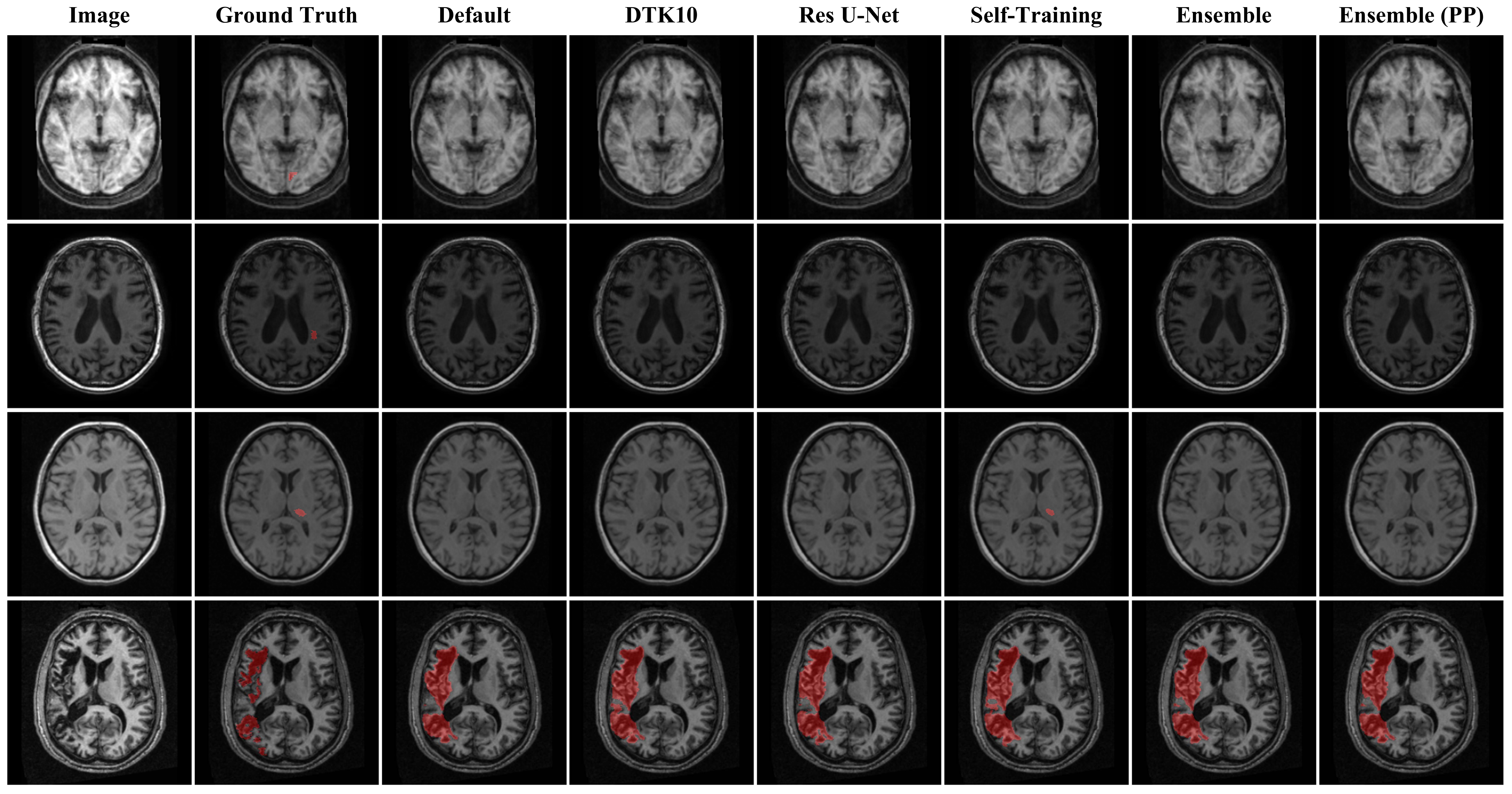}
\caption{Qualitative analysis of four failure cases. The top three rows show small lesions where the different training schemes are unable to segmentation foreground voxels due to image artifacts and poor image contrast between foreground and background. The bottom row shows several unconnected lesion regions, where the different training schemes are oversegmenting the lesion to connect the regions. } 
\label{fig:failure_cases}
\end{figure}

We also show four failure cases in Fig. \ref{fig:failure_cases}, where the different training schemes are unable to segment the lesion area accurately, or even give foreground predictions. The top three cases in Fig. \ref{fig:failure_cases} indicate that our models still can not segment small stroke lesions well in some scenarios, especially where the scans are with artifacts (the first row), or the lesion intensity is similar with surrounding areas (the second and third row). Finally, the bottom case reveals that all training schemes tend to predict unconnected lesions as a continuous lesion, especially confusing grey matter and lesion areas due to the similarities in their intensities.

\section{Discussion}
Accurately segmenting stroke lesion areas in a multi-center brain MRI dataset using an automatic segmentation method is quite challenging due to the diversity in the dataset, both in terms of scan quality and lesion size. To this end, we developed four different schemes to build diverse segmentation models, and then calculate an ensemble prediction to give better results. Additionally, we then apply an effective post-processing strategy to refine the ensemble prediction and improve some performance metrics. Experimental results on ATLAS v2.0 dataset show that our method can segment both small and big stroke lesions precisely. Our method achieve the first place on the unseen test set of the ATLAS-MICCAI challenge, which proves its generalizability and robustness. However, as can be found from the qualitative results (Fig. \ref{fig:failure_cases}), our method still can not output good predictions on some cases, especially on very small lesions or lesions with several unconnected components. Future studies will focus on improving the segmentation efficacy on small and sporadic stroke lesions. One potential solution is using generative models to create small stroke lesions to increase the diversity of the dataset \cite{huo2022brain}. Besides, some data augmentation methods \cite{zhang2018mixup,zhang2021carvemix} are worth trying as they can improve the model robustness.

\section*{Acknowledgements}
This work was supported by Centre for Doctoral Training in Surgical and Interventional Engineering at King's College London. This work was funded in whole, or in part, by the Wellcome Trust [218380/Z/19/Z, WT203148/Z/16/Z]. For the purpose of open access, the author has applied a CC BY public copyright licence to any Author Accepted Manuscript version arising from this submission. This work was supported by the UK Research and Innovation London Medical Imaging \& Artificial Intelligence Centre for Value Based Healthcare. The work was funded/supported by the National Institute for Health Research (NIHR) Biomedical Research Centre based at Guy's and St Thomas' NHS Foundation Trust and King's College London and supported by the NIHR Clinical Research Facility (CRF) at Guy's and St Thomas'. The views expressed are those of the author(s) and not necessarily those of the NHS, the NIHR or the Department of Health. We also highly appreciate the ATLAS organizers for holding the great challenge, and creating the publicly available dataset.

\bibliographystyle{splncs04}
\bibliography{books}

\end{document}